# Broadband radio-frequency TEM specimen holders for *in situ* biasing excitation and 100-ps-scale electron beam control

Christophe Gatel[1*], Marine Resano[1], Bérangère Disic[1], Nasibeh. Parsaei[2], Emmanuel Perrin[2], Hervé Parvery[2], Romain Queslier[3], Bertrand Gatti[1], Pierre Abeilhou[1], Alexandre Alibert[1], Christian Pertel[1], Cécile Marcelot[1] and Nicolas Biziere[1*]

[1] Université de Toulouse, CNRS, CEMES, Toulouse, France

[2] CISTEME, Limoges, France

[3] A3D design, Feytiat, France

*Corresponding author:
christophe.gatel@cemes.fr
nicolas.biziere@cemes.fr

## Abstract

Broadband electrical excitation inside a transmission electron microscope (TEM) is required to investigate electrically driven nanodevices at their actual operating frequencies and to develop fully electrical time-resolved experiments at a sub-nanosecond range. Here, we report the performances of homemade two- and four-channel radio-frequency (RF) TEM specimen holders based on a hybrid architecture combining coaxial transmission lines with miniaturized impedance-controlled printed circuit boards (PCB). Removable mechanical contacts provide direct electrical connection to interchangeable TEM chips without wire bonding. Vector network analyzer measurements demonstrate broadband transmission through the complete electrical path up to the specimen. With a representative TEM chip, the two-channel holder exhibits a −3 dB bandwidth of 7.9 GHz, while insertion losses remain below 12 dB up to 18 GHz.

The same RF-holder architecture was used to drive a miniaturized electrostatic electron-beam deflector. The measured temporal response indicates that the holder–deflector assembly remains compatible with electrical pulse durations approaching 100 ps, with the shortest accessible times presently limited by the available waveform generator. High-resolution TEM imaging with a 200 ps probe window does not show detectable degradation of spatial resolution, while off-axis electron holography is successfully performed for probe durations down to 0.5 ns under conditions representative of *operando* measurements.

These results establish a versatile broadband platform combining GHz electrical excitation and sub-nanosecond electron-beam control. The dual-stage configuration of our Hitachi HF3300 microscope further provides the basis for a future fully electrical pump–probe TEM approach using independently driven RF holders for specimen excitation and electron-probe control.

## Keywords

Radio-frequency TEM, specimen holder, *in situ* TEM, electron beam deflection, time-resolved TEM, off-axis electron holography

## Highlights

- Broadband RF TEM holders delivering GHz signals directly to the specimen.
- Two- and four-channel holders using removable TEM chips without wire bonding.
- RF holder–deflector response compatible with ~100 ps pulse durations.
- High-resolution TEM preserved down to 200 ps electron-probe window.
- Off-axis electron holography down to 0.5 ns probe windows.

## Introduction

Transmission electron microscopy (TEM) has progressively evolved from a technique primarily dedicated to static structural characterization towards an experimental platform capable of probing materials and devices under operating conditions. *In situ* electrical biasing, heating, mechanical loading, gas or liquid environments, and optical excitation can now be combined with high-resolution imaging, diffraction and spectroscopic techniques [1,2]. *In situ* electrical excitation is particularly relevant for functional nanodevices, as it allows to investigate their response under conditions close to those encountered during operation [3–8]. However, compared with conventional direct-current (DC) or relatively low-frequency electrical biasing, radio frequencies (RF) and microwave excitation remains comparatively uncommon in *in situ* TEM [9–11] due to instrumental constraints despite the importance of GHz dynamics in many nanoscale functional devices [12,13].

This limitation is particularly relevant in nanomagnetism and spintronics. Characteristic magnetic dynamics, including magnetization precession, vortex-core motion, and spin-wave excitation and propagation, commonly occur from hundreds of MHz to several GHz.[13–16] Access to this frequency range inside a TEM makes it possible to combine nanoscale magnetic imaging with excitation at the actual operating frequencies of the devices. Recent experiments coupling GHz electrical excitation of the sample with stroboscopic Lorentz TEM have, for example, enabled direct visualization of spin-wave generation, propagation, reflection and interference [17], illustrating the considerable potential of time-resolved electron microscopy for investigating high-frequency magnetic phenomena.

Accessing such dynamics calls for temporal resolution well beyond that of conventional TEM acquisition. Laser-driven ultrafast electron microscopy (UTEM), based on the generation of short electron packets synchronized with optical excitation of the specimen, has achieved temporal resolutions extending from the picosecond into the femtosecond regime and has enabled the observation of a broad range of ultrafast structural, electronic and magnetic phenomena [18–24]. These approaches nevertheless rely on pulsed laser systems together with dedicated electron source and optical infrastructure. Furthermore, for devices whose operation is intrinsically electrical, electrical excitation constitutes a natural and direct pump mechanism and avoids the need to reproduce an electrically driven phenomenon through optical excitation.

Laser-free schemes based on RF modulation of a continuous electron beam offer an alternative route to time-resolved TEM [25–28]. In these approaches, dedicated RF deflection elements periodically select short electron packets from the continuous beam, with temporal resolutions in the few-tens-of-picoseconds range and values down to approximately 10 ps demonstrated experimentally. Such performance makes RF-driven electron pulsing a powerful route towards laser-free stroboscopic TEM. In the implementations reported to date, however, this approach has required a dedicated pulsing assembly integrated into the microscope column, together with substantial mechanical modification and precise electron-optical alignment.

Preserving the electron-optical properties of the probe is also essential when extending these techniques beyond conventional TEM imaging. RF beam sweeping has been shown to induce an anisotropic reduction of the electron-beam spatial coherence, as evidenced by direction-dependent changes in Fresnel-fringe visibility [27,28], which may affect phase-sensitive imaging techniques such as electron holography[29–31]. To our knowledge, off-axis electron holography has not yet been experimentally demonstrated using an RF-pulsed beam.

However, generating a short electron probe addresses only one side of an electrically driven pump–probe TEM experiment. To exploit a temporal resolution of a few tens of picoseconds, the electrical excitation must also reach the specimen with an appropriate bandwidth. This differs fundamentally from demonstrating excitation at a particular microwave frequency. For continuous-wave excitation at a single frequency, propagation losses can partly be compensated by increasing the input RF power. In contrast, delivering nanosecond or sub-nanosecond electrical pulses with a well-defined waveform involves the transmission of a broad frequency range. Frequency-dependent attenuation, impedance mismatches and resonances can then distort the pulsed signal reaching the specimen, degrading both its rise time and temporal profile [32]. Broadband electrical access to the specimen is therefore a necessary counterpart to the development of an ultrafast, sub-nanosecond electron probe.

Compared with electron beam pulsing, broadband RF delivery to the TEM specimen has received comparatively much less attention. Published implementations have generally relied on custom-built holders developed for a specific experiment, for which the broadband transmission characteristics of the complete electrical path are only partially documented [9,26,28]. Operation at frequencies extending into the GHz range has been demonstrated, but the broadband insertion loss of the complete electrical path to the specimen is rarely measured. For example, RF excitation has been performed while sweeping frequencies up to 18 GHz [9], and the authors reported increased losses above 10 GHz, inferred from the reduction of the electron-beam deflection, but did not quantify the insertion loss of the complete transmission path. A single-frequency sinusoidal excitation can still be used in the presence of significant attenuation if sufficient RF power is available. In contrast, faithfully preserving the rise time and temporal profile of a fast electrical waveform requires broadband transmission of its frequency components. Consequently, the highest frequency at which an RF-induced response can be detected should not be confused with the usable bandwidth of the specimen holder.

The importance of this limitation has been directly illustrated in electrically driven stroboscopic TEM experiments. In an early implementation [26], insertion losses through a conventional biasing holder were approximately 20 dB at 2.6 GHz and approached 50 dB at 4.2 and 5.6 GHz, owing to the succession of coaxial connections, bare wires, PCB sections and wire bonds between the RF source and the specimen. A subsequent dedicated RF specimen holder substantially improved transmission over the 0–6 GHz range [27] and was later exploited for GHz stroboscopic experiments, including measurements performed at 5.25 GHz [11,17]. This represented a significant advance in RF delivery to the specimen, but the holder architecture remained based on a single RF transmission path, with the TEM specimen connected to the PCB by wire bonding and a 50 Ω termination used for impedance matching.

Here, we report broadband TEM specimen holders incorporating either two or four independent RF channels. The electrical path combines vacuum-compatible coaxial transmission lines with miniaturized impedance-controlled printed circuit boards and was experimentally characterized by S-parameter measurements over the full 0–20 GHz range. With the two-channel holder equipped with a $Si/Si_3N_4$ chip on which conducting Au tracks were lithographed, representative of those intended for TEM experiments, a measured −3 dB bandwidth of 7.9 GHz was obtained, while insertion losses remained below 12 dB up to 18 GHz. Direct removable mechanical contacts between the holder and the TEM chip eliminate wire bonding, enabling rapid specimen exchange while maintaining broadband electrical access.

The availability of several independent RF channels considerably extends the range of possible experimental configurations compared with a single signal line referenced to ground. The two- and four-channel architectures can accommodate a variety of RF excitation schemes, including single-ended and differential configurations as well as multi-terminal devices. Even the two-channel architecture allows independent or differential driving of two electrodes, as exploited here for a miniaturized ultrafast electrostatic beam deflector, whereas the four-channel version extends this flexibility to more complex multi-electrode devices.

The RF holders therefore go beyond GHz continuous-wave excitation of a TEM specimen. Their broadband response also allows fast electrical waveforms to be transmitted and the same holder technology to be used for rapid beam deflection. Taking advantage of the specific configuration of our Hitachi HF3300 microscope (I2TEM) [33,34], in which two specimen holders can be inserted simultaneously at different positions along the electron-optical axis, these two functions can ultimately be combined: an upstream RF holder is used to deflect the electron beam, while a second holder independently delivers a synchronized electrical excitation to the specimen. We demonstrate fast beam deflection with a temporal resolution presently limited primarily by the available waveform generator to approximately 100 ps, while retaining compatibility with high-resolution TEM and off-axis electron holography. The resulting architecture thus constitutes the building blocks for a versatile, fully electrical platform combining broadband *in situ* RF excitation and stroboscopic time-resolved TEM.

## 1. Design of the broadband RF TEM specimen holders

### 1.1. Design constraints and hybrid RF architecture

Integrating broadband RF transmission lines into a TEM specimen holder involves constraints that are considerably more stringent than those encountered in conventional microwave circuits. In addition to the very limited space available within the holder body and around the specimen region, all components must be compatible with high vacuum, exhibit very low outgassing, and remain non-magnetic. At the same time, the RF path must maintain a characteristic impedance close to 50 Ω while minimizing insertion losses and reflections over a broad frequency range. Mechanical, vacuum and electromagnetic constraints were therefore considered simultaneously throughout the design process.

The geometry of the I2TEM specimen stages imposes an additional constraint on the RF-holder design. In the Hitachi side-entry goniometers, the holder extends far beyond the specimen position and terminates by a sapphire tip forming part of the mechanical support and guidance within the stage. This geometry differs from other TEM platforms in which a relatively large PCB carrier can be mounted directly at the holder tip. In the I2TEM configuration, the specimen region is instead surrounded by the mechanical structure of the holder, leaving only a few millimeters for integrating the RF lines and the interchangeable TEM chip.

A fully coaxial architecture would offer the lowest propagation losses but cannot be extended up to the specimen due to geometrical constraints of the holders. Conversely, routing the RF signal over the entire holder length using a PCB would result in excessive high-frequency losses over a distance of approximately 36 cm. A hybrid architecture was therefore adopted, combining coaxial transmission lines over most of the holder length with miniaturized PCB transmission lines only in the confined region close to the specimen (Figure 1a).

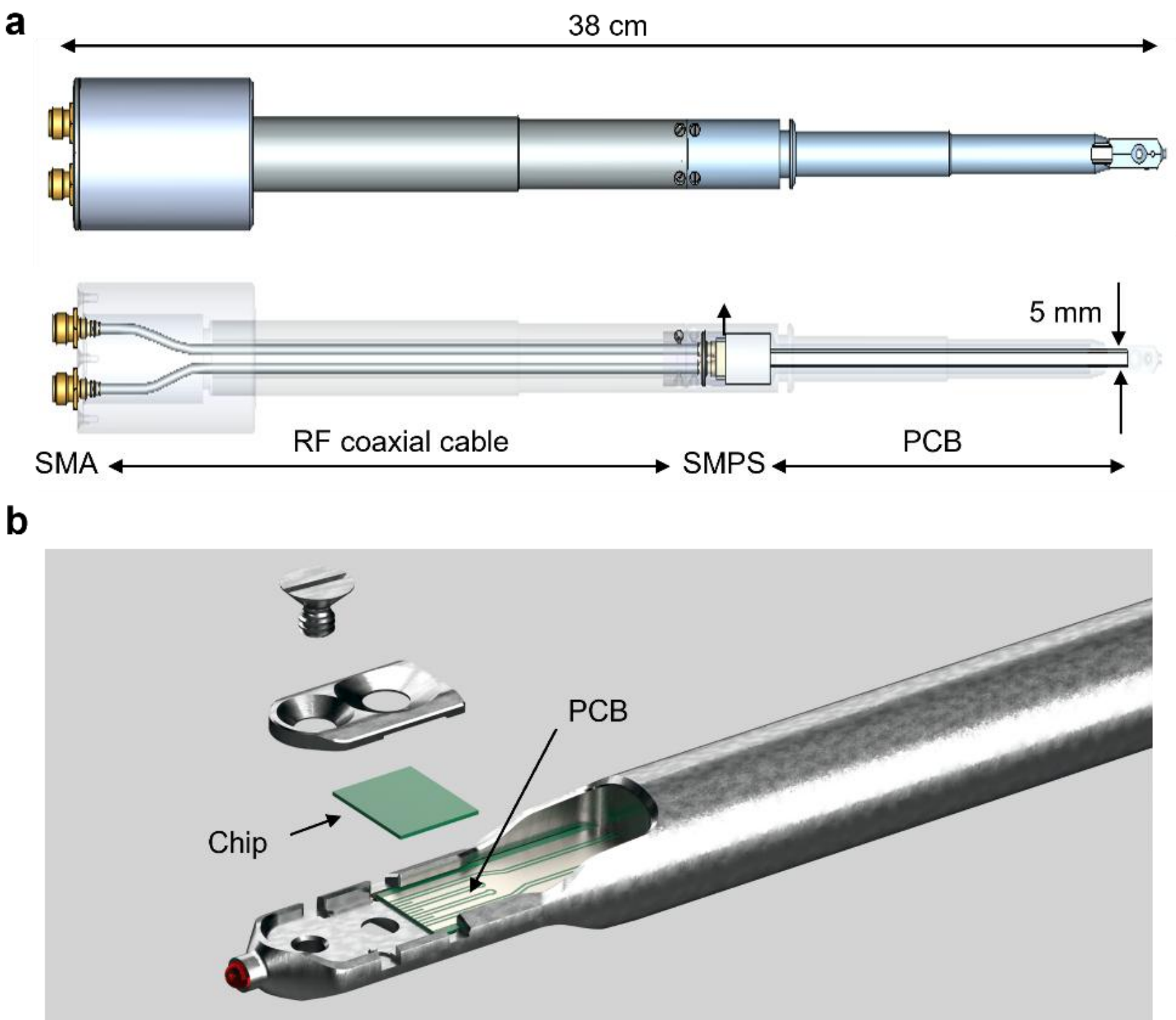

***Figure 1.*** *(a) Schematic representation of the RF TEM specimen holder showing the hybrid transmission architecture. RF signals are transferred from the external SMA connectors through non-magnetic RF coaxial cables and subsequently to miniaturized PCB transmission lines through SMPS-SV transitions near the holder head. (b) Detail of the holder–chip interface. The interchangeable TEM chip is mounted face-down onto the RF contact tracks and mechanically pressed and aligned by a removable metallic cover, providing direct electrical contact without wire bonding.*

At the external end of the holder, standard SMA connectors form the interface with the RF instrumentation. Fully non-magnetic RF coaxial cables carry the signals along the holder body to the vicinity of the PCB section. The transition towards the PCB section is achieved using compact SMPS-SV connectors, selected to satisfy the severe spatial constraints while preserving the microwave properties of the transmission path. The PCB section then routes the signals to the holder head through coplanar waveguides (CPWs) in a ground–signal–ground (GSG) configuration. This geometry confines the electromagnetic field around the transmission line and limits radiation losses.

The PCB substrate had to satisfy both microwave and vacuum requirements. The restricted lateral dimensions of the holder impose narrow transmission lines and favors a relatively high-permittivity dielectric substrate for maintaining RF capabilities. Although ceramic materials such as alumina are particularly well suited to vacuum environments, Rogers RO3010 was selected as a practical compromise between dielectric properties, low outgassing, manufacturability, mechanical stability and fabrication cost.

The resulting architecture therefore uses the lowest-loss coaxial transmission medium over the longest possible part of the electrical path and restricts the PCB section to the region where insertion of two or more coaxial cables becomes unavoidable. The use of coplanar waveguides also maintains controlled signal and ground conductors up to the vicinity of the specimen, rather than relying on isolated electrical leads or on the holder body itself as the sole electrical ground reference.

A second critical aspect of the design concerns the electrical interface between the holder and the TEM specimen chip. Instead of wire bonding, which complicates specimen exchange, and may gradually deteriorates the electrical pads and be particularly problematic for fragile electron-transparent devices, a fully removable mechanical contact was developed (Figure 1b). The chip is mounted face-down onto the RF contacts located at the holder head, and a dedicated cover applies mechanical pressure while ensuring alignment, planarity and stability of the electrical interface. Samples can therefore be exchanged without wire bonding or additional microfabrication steps after insertion onto the holder.

Finally, both holders were designed as a platform that can be easily and completely disassembled dismantled to allow for repairs or modifications to the internal architecture, for example by replacing the PCB to change the number or dimensions of the lines.

### 1.2. Two- and four-channel configurations and removable chip interface

Two versions of the RF holder were developed, providing respectively two and four independent RF channels. In the two-channel configuration, two CPW transmission lines connect the SMPS interfaces to the holder head (Figure 2a). The four-channel version in Figure 2b required a more complex PCB architecture because four RF transmission lines could not be accommodated on a single surface within the available holder width. Indeed, the limited lateral dimensions of the PCB do not allow the GSG geometry to be optimized for a 50 Ω characteristic impedance, particularly with respect to the spacing between the signal and ground conductors. A second pair of RF lines was therefore routed on the opposite side of the main PCB. The additional channels are transferred between the bottom and top PCB surfaces through metallized vias. Since the limited thickness of the main board does not allow SMPS connectors to be mounted directly on both faces with sufficient vertical separation, two additional transition PCBs were introduced on opposite sides of the main board to carry the connectors. This arrangement preserves the required connector spacing while routing all four RF channels towards the holder head.

The line geometries and transitions were optimized to maintain an impedance close to 50 Ω throughout the RF path while accounting for the actual mechanical constraints of the assembled holder. In the four-channel version, however, the restricted lateral space available for two adjacent GSG transmission structures inevitably limits perfect impedance matching over the entire frequency range.

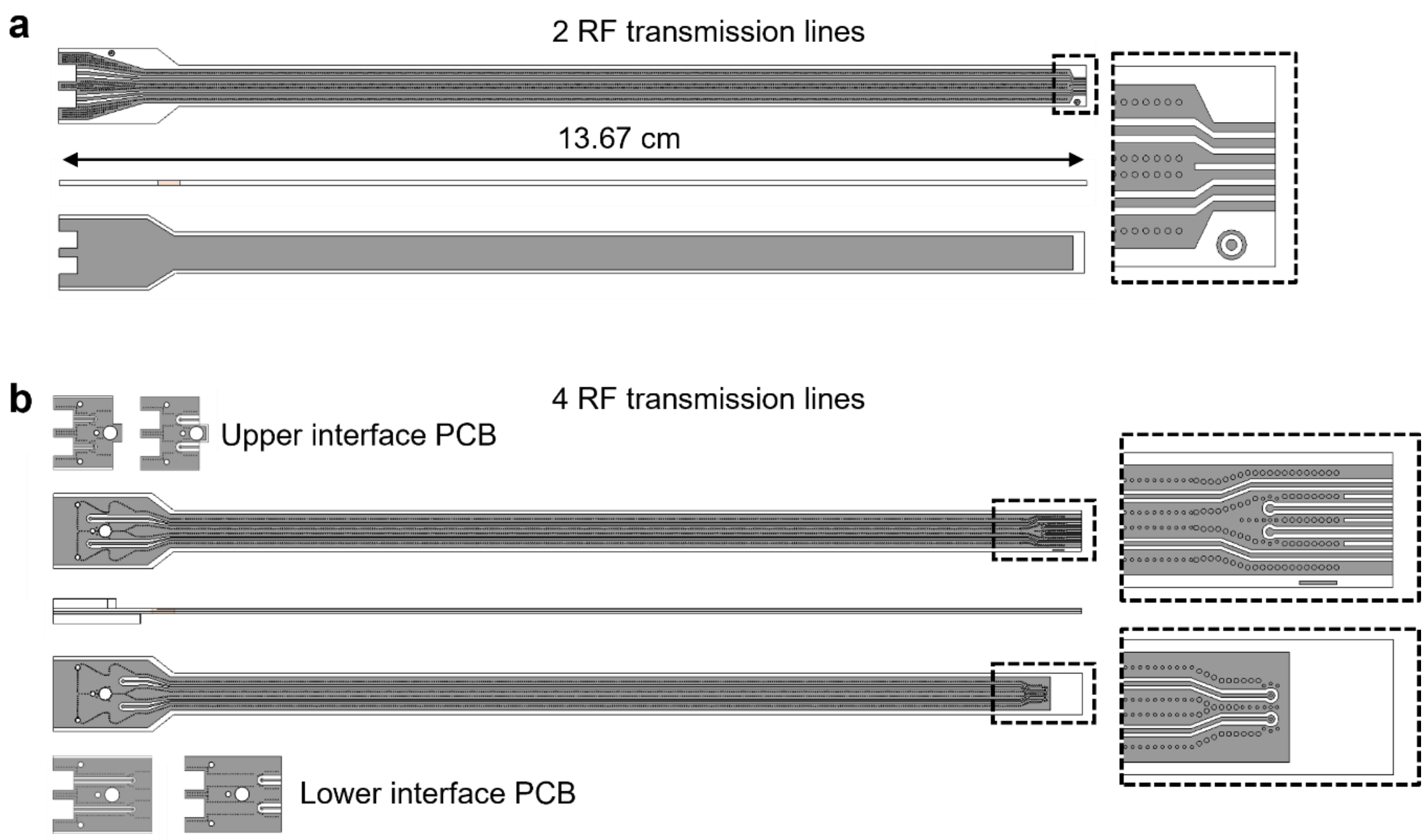

***Figure 2.*** *(a) Two RF transmission lines based on a two-layer RO3010 PCB carrying two coplanar waveguides in ground–signal–ground geometry between the SMPS-SV transitions and the holder head. (b) Four RF transmission lines. A second pair of RF lines is routed on the opposite side of the main PCB, with metallized vias and two auxiliary transition PCBs allowing the SMPS connectors to be accommodated on both sides while maintaining the required mechanical dimensions and an impedance close to 50 Ω.*

The specimen geometry is defined so that the electrical tracks patterned on the TEM chip coincide reproducibly with the RF contacts at the holder head while the electron-transparent membrane remains centered on the optical axis. At the holder–chip interface, the width of the RF tracks depends on the holder version and are between 160–250 μm wide, with comparable inter-track spacings. The metallization patterned on the interchangeable TEM chips must therefore reproduce the corresponding track widths, spacing and lateral positions to ensure reliable electrical contact while avoiding short circuits between adjacent channels. In the present design, the chips have nominal dimensions of 5.1 $\pm0.02 \times 6.05 \pm0.02$ mm$^2$, with a $Si_3N_4$ membrane positioned 3 mm from the lower edge. These dimensional tolerances are particularly relevant to the RF interface, since lateral displacement of the chip or of the patterned contact tracks by only a few tens of micrometers already represents a significant fraction of the conductor width or gap. Nevertheless, track dimensions and alignment tolerances in this range are readily achievable using standard cleanroom microfabrication processes and do not constitute a significant limitation for routine chip fabrication.

The removable-chip concept is central to the versatility of the holders. The RF transmission architecture is integrated into the holder, whereas the final electrical configuration is defined by the metallization patterned on the interchangeable TEM chip. The same holder can therefore accommodate a variety of single-ended, differential or multi-terminal excitation schemes. Two independent channels enable a broad range of excitation schemes, including independent or differential driving of two electrodes of the electrostatic beam deflector described below, while the four-channel architecture extends this flexibility to more complex multi-electrode devices.

## 2. RF characterization of the specimen holders

### 2.1. Broadband characterization using dedicated RF test chips

The RF performances of both holder configurations were characterized using a ZNA43 Rohde et Schwarz vector network analyzer (VNA) over the frequency range from 10 MHz to 20 GHz. Because direct access to the RF lines at the holder head is not possible once the holder is assembled, dedicated PCB test chips were designed to create a through-transmission path between two channels. These test chips incorporate impedance-matched coplanar transmission lines electrically connecting two RF channels at the specimen position, with their track dimensions adapted to those of the corresponding RF contacts at the holder head (Figure 3a).

For the two-channel holder, transmission measurements were performed from port 1 to port 2 ($S_{21}$) through the dedicated test chip to characterize both transmission channel, each including the SMA connector, one RF cable, the coaxial to PCB transition and one PCB to chip transition. The measured $S_{21}$ signal then allows to obtain the insertion losses of each individual RF channel by dividing the signal by two, assuming equivalent losses in the both channels.

The same procedure was applied to the four-channel holder using the channel pairs 1–4 and 2–3. This measurement configuration includes the successive coaxial, coaxial-to-PCB and PCB-to-chip transitions encountered by the RF signal along the complete transmission path. The extracted insertion losses therefore characterize the assembled RF system rather than the propagation loss of the PCB transmission lines alone.

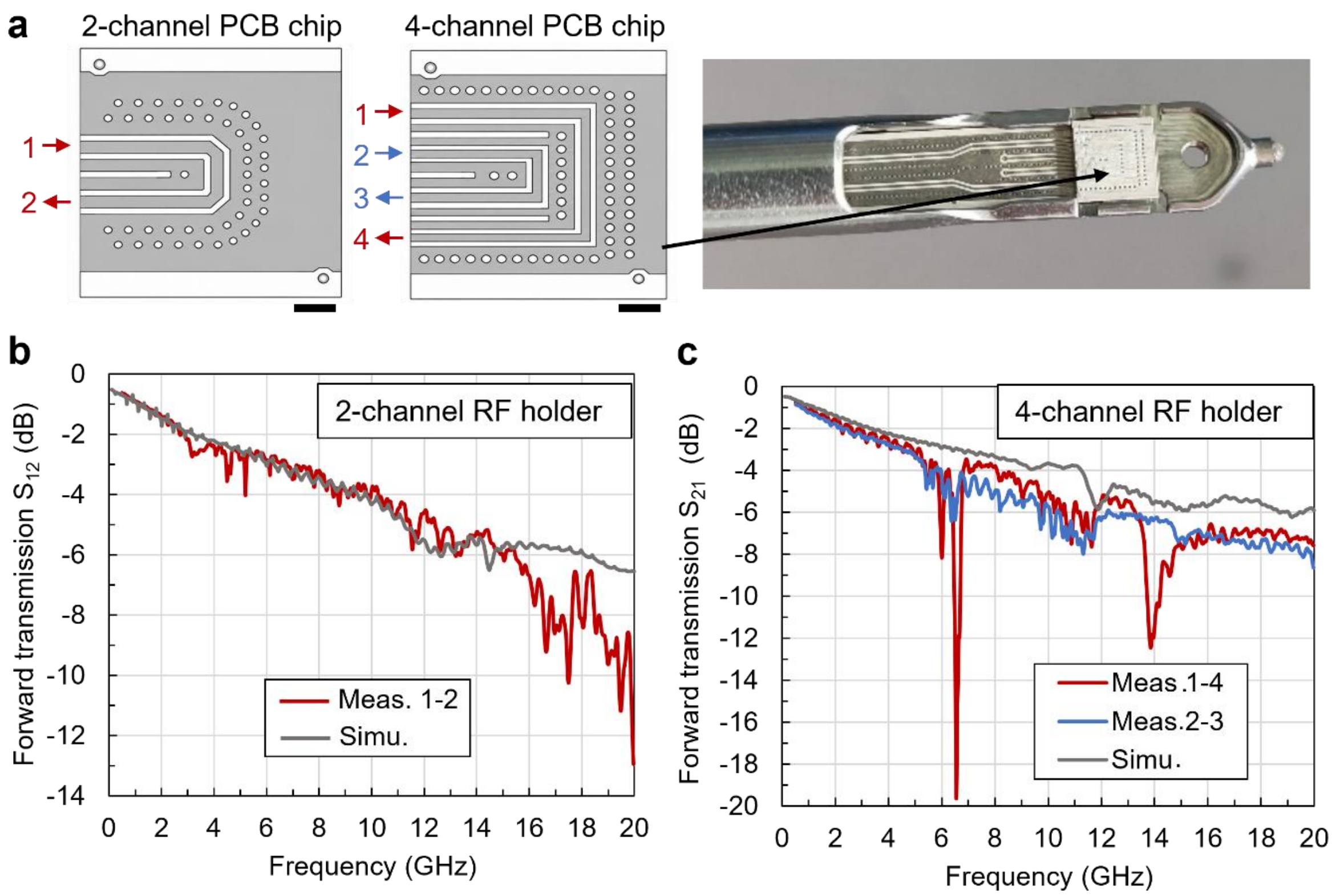

***Figure 3.*** *(a) Geometry of the transmission measurement and dedicated impedance-matched PCB test chips used to connect two RF channels at the specimen position. For the two-channel holder, transmission is measured between channels 1 and 2; for the four-channel holder, the 1–4 and 2–3 channel pairs are characterized. The attenuation reported for an individual RF path corresponds to half of the $S_{21}$ measurements. (b) Measured insertion losses as a function of frequency for the two-channel holder with the corresponding simulated response. (c) Measured insertion losses as a function of frequency for the four-channel holder with the corresponding simulated response. Measurements were performed using a VNA from 10 MHz to 20 GHz. Scales bars in (a) are 1 mm.*

The measured responses were compared with the RF simulations performed during the design stage (Simulia CST Studio) as shown Figures 3b and 3c. For the two-channel configuration, the insertion loss remains below 3 dB up to approximately 6 GHz and below 10 dB up to approximately 18 GHz (Figure 3b). For the four-channel configuration, the −3 dB bandwidth reaches approximately 4.5 GHz, while insertion losses remain below approximately 9 dB up to 20 GHz, apart from localized loss peaks observed for the 1–4 transmission path at approximatively 6.5 GHz and 14 GHz (Figure 3c).

The differences between simulations and measurements, including localized loss peaks, can partly be attributed to the holder-to-chip interface used during these characterization experiments. In this test configuration, the dedicated PCB chips were clipped onto the holder head rather than being pressed and accurately positioned by the metallic cover used under normal operating conditions (Figure 3a). Contact pressure and alignment were therefore less reproducible, introducing additional losses that are not intrinsic to the RF transmission lines of the holders.

The bandwidths measured with these dedicated test chips should consequently be regarded as conservative estimates of the usable RF performance of the holders under these specific contacting conditions. Since part of the measured attenuation originates from the non-optimal test-chip interface, the intrinsic transmission performance of the holders is expected to be at least comparable, and likely better, than that measured in this configuration.

### 2.2. RF performance with a TEM-compatible Si/$Si_3N_4$ chip

To evaluate the RF response under conditions representative of an actual TEM experiment, a dedicated Si/$Si_3N_4$ chip was fabricated for the two-channel holder (Figure 4a). Its dimensions and contact geometry reproduce those of the interchangeable TEM chips described in Section 2. The chip contains a 100×100 μm² electron-transparent $Si_3N_4$ window and Au conducting tracks electrically connecting channels 1 and 2, allowing through-transmission measurements to be performed with a substrate and metallization geometry representative of those intended for *in situ* TEM experiments.

The corresponding transmission response, measured here using a MS4642A Anritsu VNA, is compared in Figure 4b with that obtained using the dedicated PCB test chip. The two measurements exhibit a similar overall frequency dependence. Interestingly, the losses measured with the Si/$Si_3N_4$ chip are slightly lower up to approximately 12 GHz, and the −3 dB bandwidth extends to 7.9 GHz, following by a localized attenuation peak slightly above 8 GHz. This peak may be caused by a cavity-like or standing-wave effect due to the small impedance mismatch at the PCB-chip interface, combined with the specific length of the coplanar waveguides on the chip, leading to partial destructive interference with the propagating signal. Above approximately 12 GHz, the attenuation becomes slightly larger than that measured with the PCB test chip. This may be attributed, at least in part, to the limited thickness of the metallic pads (≈ 200 nm) which increases ohmic losses compared with thicker conductors. Nevertheless, insertion losses remain below approximately 12 dB up to 18 GHz. Overall, the precise origin of the attenuation peaks and frequency-dependent response of the TEM-compatible chip likely results from the combined effects of the chip geometry, the Si substrate and associated impedance discontinuities. More importantly, the close agreement between the two types of test structure over most of the measured frequency range confirms that the broadband RF performance obtained with the dedicated PCB structures is largely preserved when a realistic TEM chip is mounted on the holder.

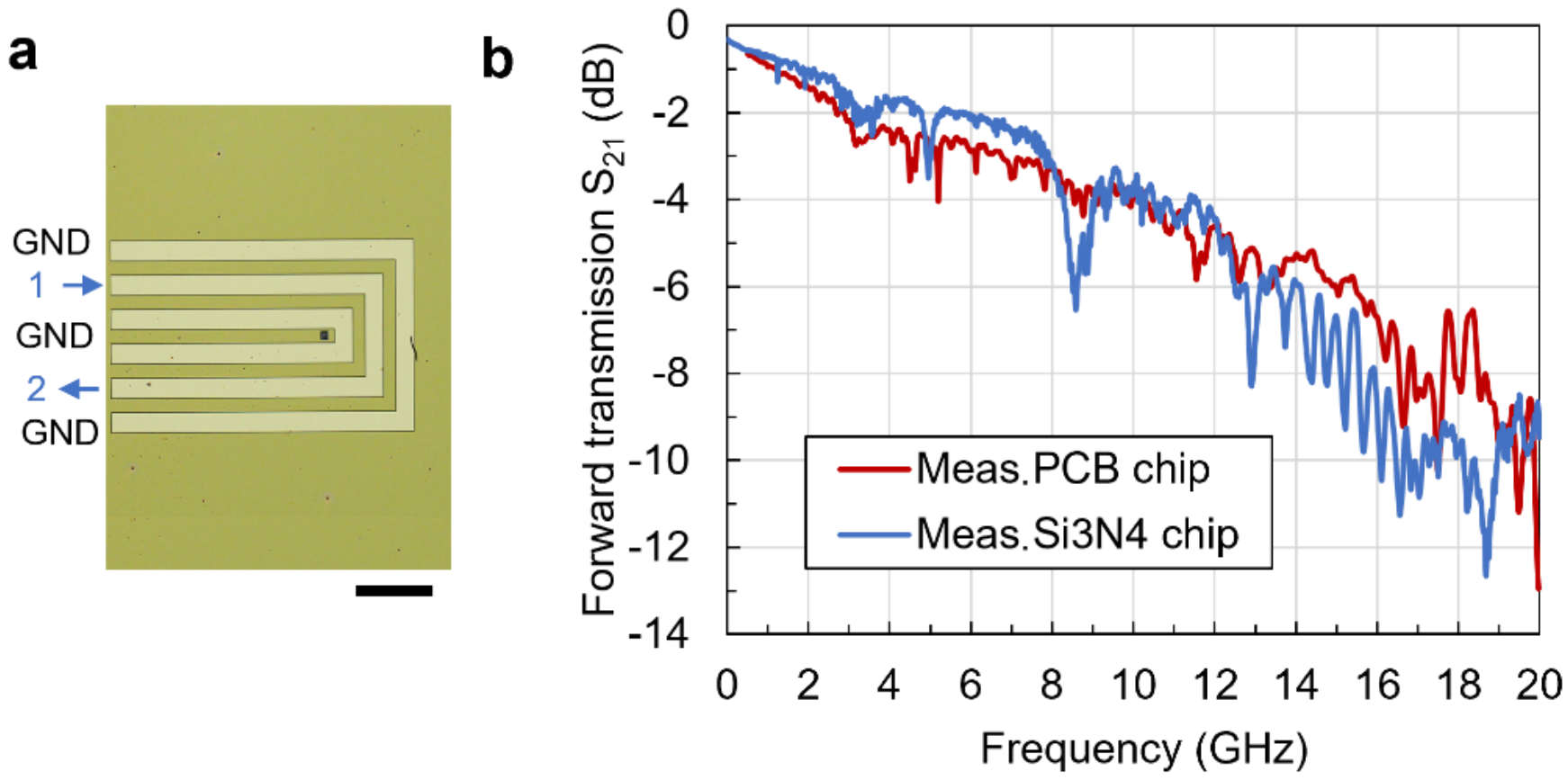

***Figure 4.*** *(a) Design and photograph of the Si/$Si_3N_4$ test chip used with the two-channel RF holder. Au tracks connect channels 1 and 2 and surround a 100×100 μm² electron-transparent $Si_3N_4$ window (black square in the middle part) representative of the chips intended for in situ TEM experiments. (b) Insertion loss measured with the Si/$Si_3N_4$ chip compared with that obtained using the dedicated PCB test chip. The Si/$Si_3N_4$ configuration exhibits a −3 dB bandwidth of 7.9 GHz and insertion losses below 12 dB up to 18 GHz. The localized attenuation above 8 GHz is associated with the chip geometry and Si-substrate response. Scale bar in (a) is 1 mm.*

The −3 dB bandwidth of 7.9 GHz measured in this configuration demonstrates that RF signals extending over several GHz can be transmitted through the complete holder and delivered directly to the TEM specimen with limited attenuation. This distinction is important: the measured performance corresponds to electrical transmission up to the sample itself, under conditions representative of TEM operation, rather than simply to the transmission properties of the holder terminated before the specimen interface.

### 2.3. Implications for fast electrical waveforms

As discussed in the Introduction, broadband transmission becomes particularly important when the applied excitation is a pulse or another rapidly varying waveform. A sinusoidal excitation at a single frequency may remain usable even in the presence of substantial attenuation if sufficient RF power is available. In contrast, preserving the rise time and temporal profile of a fast electrical waveform depends on the simultaneous transmission of a broad spectrum of frequency components. Strongly frequency-dependent attenuation or resonances would therefore directly distort the excitation reaching the specimen.

As a first-order estimate, the relationship between the −3 dB bandwidth $BW$ and the 10–90% rise time of a bandwidth-limited system,

$$t_{10-90} \approx \frac{0.35}{BW} \qquad \text{Eq. (1)}$$

gives a characteristic rise time of approximately 45 ps for the 7.9 GHz bandwidth measured with the TEM-compatible $Si/Si_3N_4$ chip on the two-channel version. The measurements on the four-channel version using the PCB test chip corresponds to a rise time lower than 80 ps. This estimate also depends on the waveform generator and the specimen under study. It nevertheless shows that the RF transmission path itself is compatible with electrical variations occurring well below the nanosecond timescale. This broadband electrical capability is exploited in the following section, where the same RF-holder architecture is used to drive a miniaturized electrostatic beam deflector.

## 3. Fast electron beam deflection for an all-electrical stroboscopic approach

### 3.1. Dual-stage concept and deflector design

A key feature of the I2TEM microscope is its double-stage configuration. An upper Lorentz stage is located above the objective lens, while a conventional specimen stage is positioned between the objective-lens pole pieces (Figure 5a). The two object planes are separated by 48 mm, and the RF holders described above can be inserted independently into either stage. Two RF holders can therefore be operated simultaneously within the microscope.

This configuration provides the basis for a future fully electrical pump–probe approach. In the configuration considered here, an RF holder inserted into the upper Lorentz stage equipped with a fast electrostatic deflector can control the temporal electron probe, while a second RF holder inserted into the conventional stage delivers an independent electrical excitation to the specimen (Figure 5b). Once the two electrical signals are synchronized, the signal applied to the specimen will define the pump, whereas the time interval $t_{view}$ during which the electron beam is directed towards the selected observation position will define the probe window.

For an excitation of period $t_{cycle}$, the duration of the probe window can be adjusted through the duty cycle of the waveform applied to the deflector. A 10% duty cycle, for example, defines a probe window $t_{view}$ corresponding to one tenth of the excitation period $t_{cycle}$ and reduces the electron intensity contributing to the selected image to approximately 10% of the continuous-beam intensity. For a given excitation period, decreasing the duty cycle therefore provides a narrower temporal window and allows the excitation cycle to be sampled more finely, at the expense of a proportional reduction in the number of collected electrons. By varying the relative delay between specimen excitation and beam deflection, different times within the excitation cycle could be probed stroboscopically.

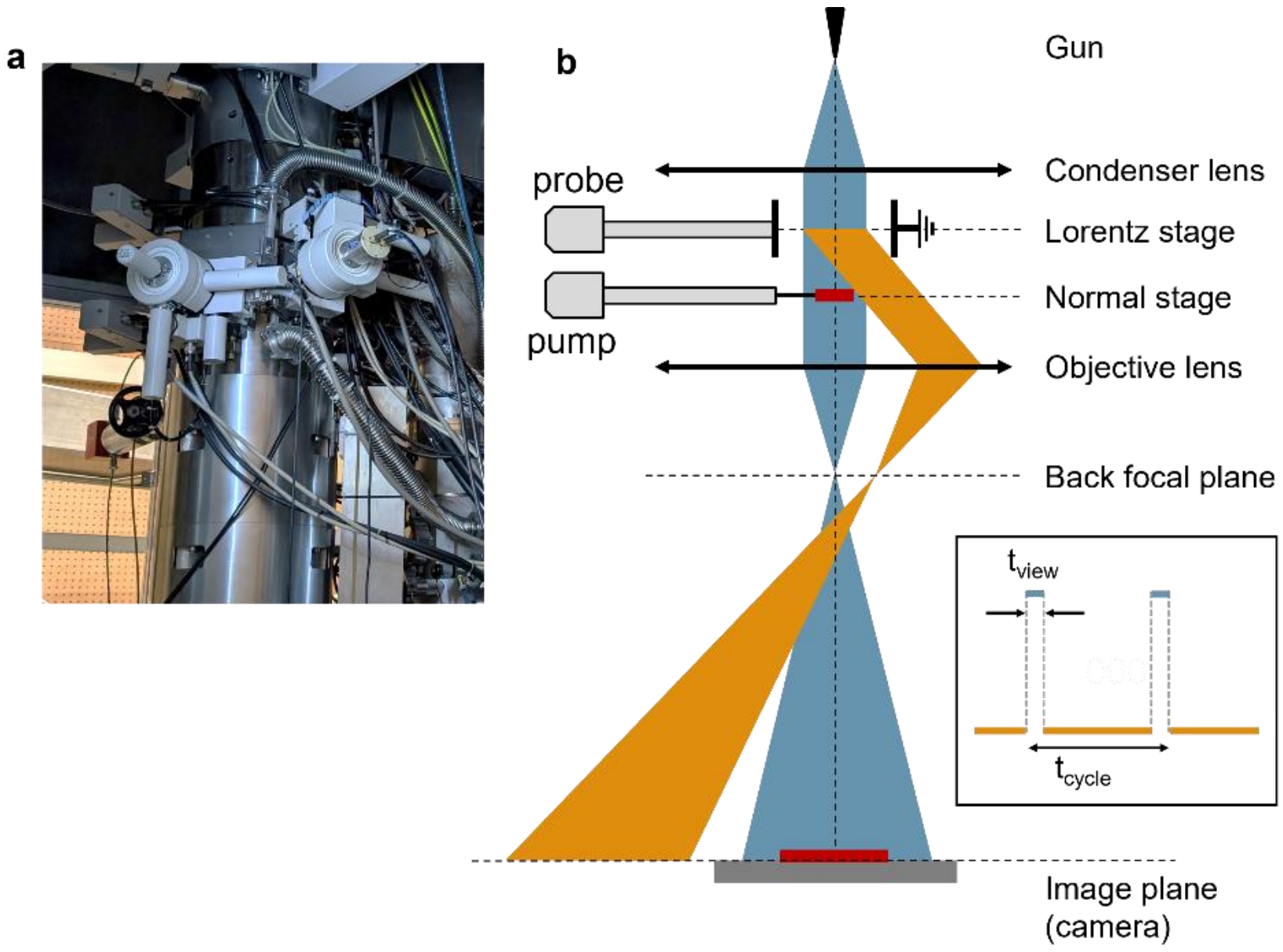

***Figure 5.*** *(a) I2TEM double-stage geometry, comprising an upper Lorentz stage and a conventional specimen stage separated by 48 mm along the electron-optical axis. An RF holder can be inserted independently into each stage. (b) Principle of the proposed fully electrical pump–probe configuration. The upper RF holder carries the electrostatic deflector and controls the electron probe, while the lower RF holder delivers an independent electrical excitation to the specimen. The probe duration is controlled by the duty cycle of the deflection waveform, and different positions within the excitation cycle can ultimately be accessed by varying the relative delay between the two synchronized electrical signals.*

This dual-holder concept motivated the development of a miniaturized electrostatic deflector directly compatible with the RF specimen holders. The deflector was designed as an interchangeable chip with the same lateral dimensions and mechanical interface as the TEM chips described in Section 1. Deflector chips were developed for both the two- and four-channel holders.

The deflector consists of two opposing electrodes separated by a narrow slit through which the electron beam propagates (Figure 6a). A voltage difference between the electrodes generates an electric field transverse to the beam direction and produces an electrostatic deflection perpendicular to the slit axis. Ground planes surrounding the electrodes are included both to preserve the RF transmission geometry and to shield the electron beam from unwanted fields originating from other regions of the PCB. The overall thickness of the deflector is 1.28 mm, as imposed by the mechanical mounting geometry of the RF holder. Deflectors with slit widths ranging from 300 to 500 μm and with different slit orientations were fabricated (Figure 6b), allowing both the amplitude and direction of the beam displacement to be adapted to the experiment. Even for the smallest 300 μm slit, the aperture remains larger than the transverse extent of the electron beam under the magnification conditions used here, so that the deflector does not geometrically restrict or clip the beam. In both holder configurations, the two deflector electrodes are connected to two independent RF channels.

The angular deflection is proportional to the voltage applied between the electrodes, and increases as the slit width decreases. After passing through the deflector, the electron beam propagates over the 48 mm distance separating the two I2TEM object planes, converting the angular deflection into a lateral displacement at the lower specimen plane.

The electrostatic response was evaluated by finite-element modelling using the actual three-dimensional geometry of the deflector as presented Figure 6c. The contribution of stray fields is particularly important in this geometry as the 1.28 mm propagation length through the deflector is only a few times larger than the 300–500 μm slit width, preventing the electric-field distribution from being accurately approximated by that of an ideal parallel-plate capacitor. The simulations show that stray fields can contribute almost 45% of the total deflection for a 300 µm slit. For the 330 µm-wide deflector subsequently used experimentally, the calculated deflection angle is 16 $\mu rad \cdot V^{-1}$, corresponding to a lateral displacement of 768 $nm \cdot V^{-1}$ at the second object plane.

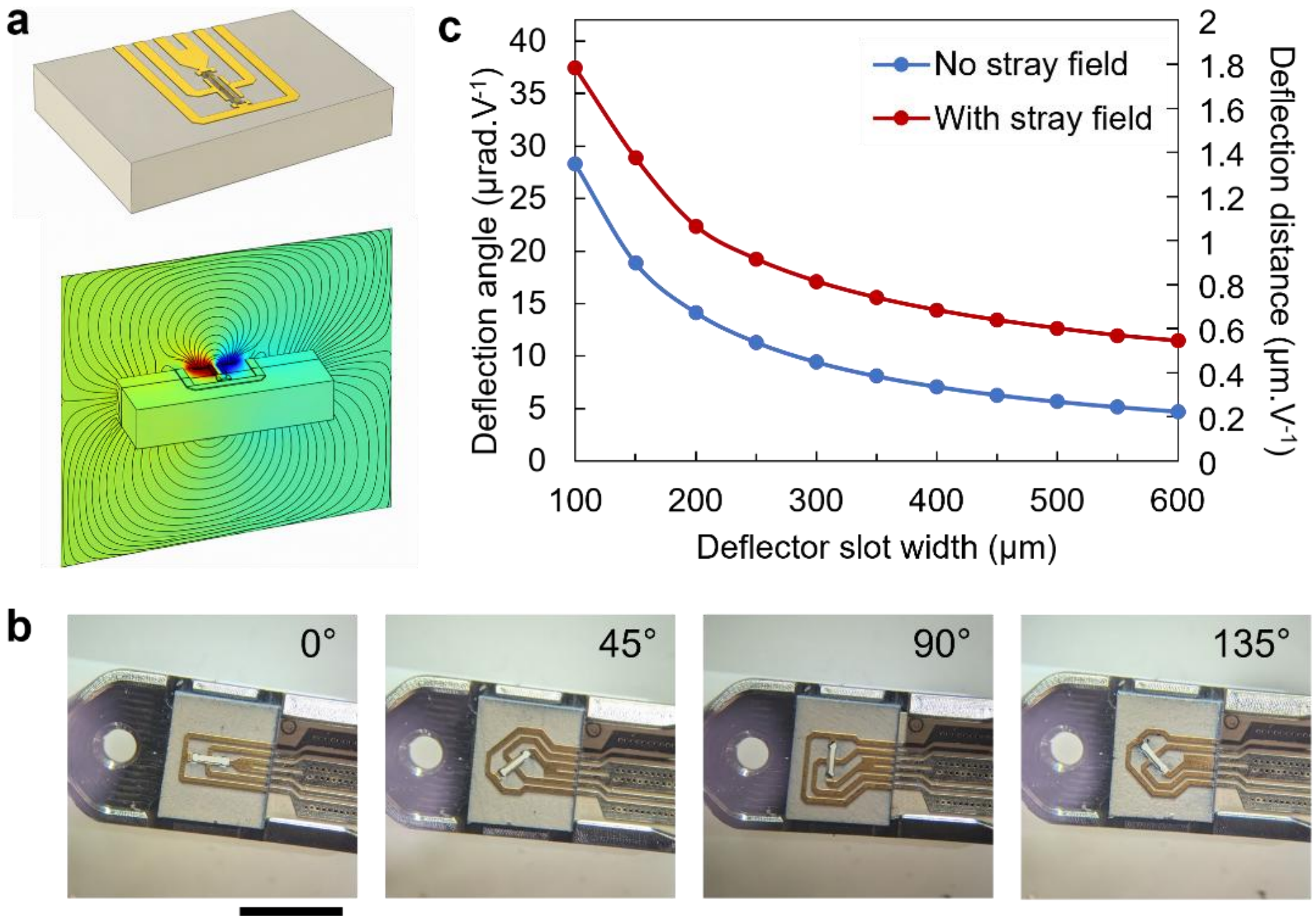

***Figure 6.*** *(a) Schematic representation of the miniaturized electrostatic deflector mounted as an interchangeable chip on the RF specimen holder. Two deflection electrodes are positioned on either side of a narrow slit through which the electron beam propagates. Surrounding ground planes preserve the RF transmission geometry and shield the beam from unwanted electric fields. The deflector thickness is 1.28 mm. Below: three-dimensional finite-element modelling (FEM) of the electric-field distribution when a voltage difference is applied between the two electrodes. (b) Examples of fabricated deflector chips with different orientations, allowing the direction of the beam deflection to be adapted to the experimental configuration. (c) FEM-calculated deflection angle per unit applied voltage as a function of slit width, with the corresponding lateral beam displacement at the second I2TEM object plane, located 48 mm downstream, shown on the right-hand axis. Calculations including and neglecting stray fields are shown for comparison. Scale bar in (b) is 5 mm.*

### 4.2. Experimental beam deflection and temporal response

Beam-deflection experiments were performed using the two-channel RF holder equipped with a 330 µm-wide deflector. The I2TEM microscope was operated in Lorentz TL11 mode and the beam was recorded using a Gatan K3 direct-electron detector in counting mode. A small condenser aperture was introduced to reduce the electron flux and avoid saturation of the detector, allowing quantitative intensity measurements. It also produces a compact electron spot, making the beam displacement easier to visualize and quantify accurately.

Fast electrical waveforms were generated using a Tektronix AWG70001B arbitrary waveform generator operated at 50 $GS \cdot s^{-1}$, with a rise time (10%-90%) of approximately 45 ps. The generator provides two complementary outputs of opposite polarity, which were connected to the two electrodes of the deflector. Each output was operated at 0.25 $V_{PP}$ into 50 Ω. The differential voltage between the two electrodes therefore varies from −0.25 to +0.25 V under matched conditions, corresponding to a total amplitude of 0.5 $V_{PP}$. The deflector, however, behaves as a high-impedance capacitive load rather than as a matched 50 Ω termination. For a 50 Ω source driving a high-impedance load, the voltage at the load approaches twice the value specified for a matched 50 Ω load. The measured output amplitude was 0.272 $V_{PP}$ per complementary output, leading to a maximum differential voltage of approximately 1.1 $V_{PP}$ across the deflector.

The deflection was first characterized using a square-wave excitation with a period of 200 ns and a 50% duty cycle. The applied voltage was progressively varied in steps of 50 mV and an electron beam image was recorded for each voltage (Figure 7). The microscope magnification was set in order to observe the two spots corresponding to the negative and positive voltage states of the square waveform for the maximum amplitude. The equal spot intensities result from the identical 100 ns residence time at each beam position. The spot displacement increases linearly with the applied voltage, as expected for the small angular deflections involved here. At the maximum voltage available from the AWG, the separation between the two beam positions reaches approximately 830 nm. Considering the maximum differential voltage of approximately 1.1 V, this corresponds to a displacement of approximatively 750 nm·$V^{-1}$, in very good agreement with the FEM prediction for a 330 μm-wide slit.

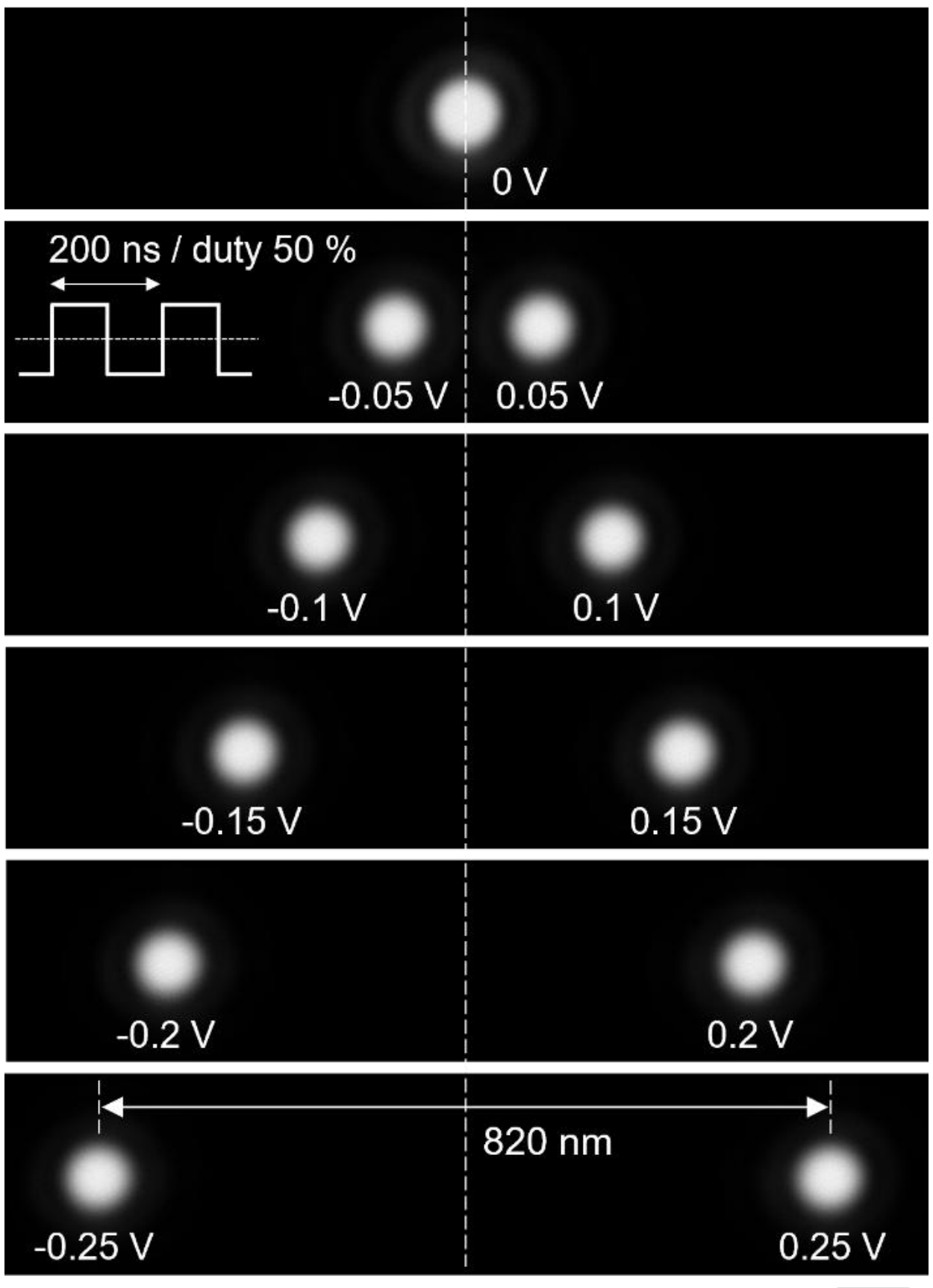

***Figure 7.*** *Electron-beam images acquired with the two-channel* RF *holder equipped with a 330 μm-wide deflector under square-wave excitation with a period of 200 ns and a 50% duty cycle. The differential voltage was progressively varied in 50 mV steps. The left and right correspond to the negative and positive voltage states of the waveform, respectively. Their separation increases linearly with applied voltage. Scale bar is 100 nm.*

The temporal behavior was then investigated by reducing the excitation period and varying the duty cycle. Figure 8a compares electron beam images acquired at the maximum generator amplitude for different waveform conditions. The two beam positions are here denoted *N* and *P*, corresponding respectively to the negative and positive voltage states applied to the deflector. For each TEM image, the corresponding electrical waveform was independently measured at the generator output using the same cables as those employed during the microscope experiments and a 13 GHz-bandwidth Agilent Infiniium oscilloscope.

For a period of 20 ns and a 50% duty cycle, the result is essentially identical to that obtained with a 200 ns period. The *N* and *P* spots remain clearly separated and exhibit the same shape, with a residence time of 10 ns at each position. The separation between the two spots, indicated by the double arrow in Figure 8a, is identical to that measured in the preceding experiment at 200 ns for the same maximum applied voltage. The electrical waveform remains close to square, with only slight rounding immediately after the transitions before reaching the full voltage amplitude over a large fraction of each pulse.

When the period is reduced to 5 ns while maintaining a 50% duty cycle, the residence time decreases to 2.5 ns at each beam position. The two spots remain perfectly distinguishable, but a weak intensity tail appears between each spot and the central region. At the same time, the electrical waveform measured at the generator output exhibits more pronounced rounding and reaches its maximum voltage less efficiently. The mean voltage over each plateau is consequently slightly lower than the maximum value obtained for longer periods, and explain why both spots remain slightly displaced from the dashed reference positions corresponding to the maximum beam deflection.

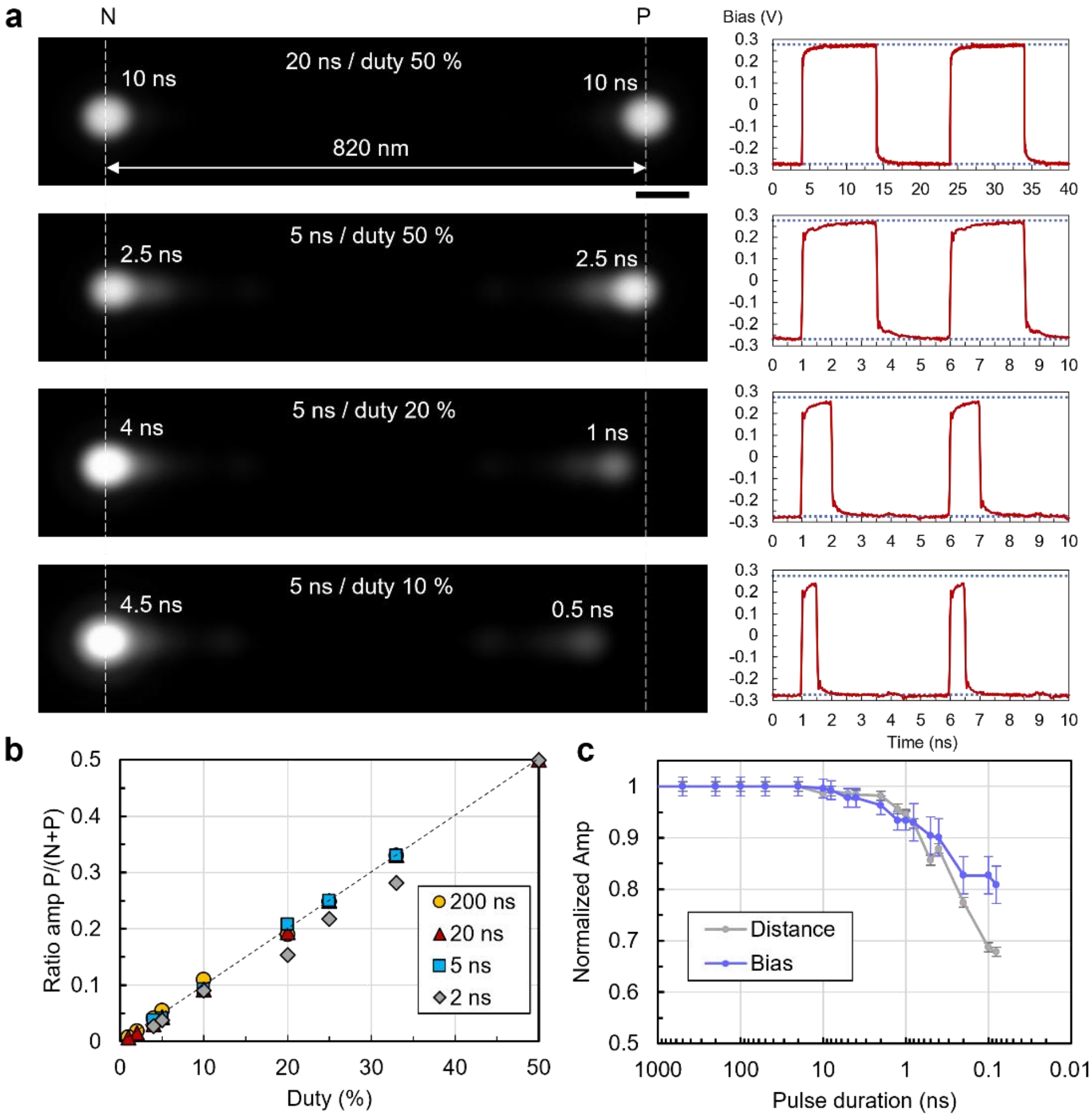

***Figure 8.*** *(a) Electron-beam deflection images obtained at the maximum AWG amplitude for different excitation periods and duty cycles, together with the corresponding electrical waveforms measured directly at the generator output using the same cables as in the TEM experiment. N and P denote the beam positions corresponding to the negative and positive voltage states, respectively. (b) Relative intensity of the P spot as a function of duty cycle for different excitation periods. The linear relationship demonstrates that the fraction of electrons collected at each beam position reflects the corresponding residence time. For several duty-cycle values, data obtained at different excitation periods overlap closely, particularly at*

*50% duty cycle. (c) Normalized beam-deflection amplitude (distance) and normalized electrical-pulse amplitude (bias) delivered by the AWG as a function of pulse duration. The two responses closely follow each other down to approximately 0.4 ns; at shorter durations, the holder–deflector response remains compatible with pulse durations approaching 100 ps, close to the operating limit of the available AWG. Scale bar in (a) is 100 nm.*

Keeping the 5 ns period while reducing the duty cycle produces an intentional imbalance between the residence times at the two beam positions. At a 20% duty cycle, the *P* spot corresponds to a residence time of 1 ns, compared with 4 ns for the *N* spot. The longer negative-voltage interval now allows the generator output to recover its full negative level, and the *N* spot consequently reaches the dashed reference position. The resulting difference in spot intensity directly reflects the unequal residence times. When the duty cycle is further reduced to 10%, the residence time of the *P* spot reaches only 0.5 ns. Despite its substantially lower intensity, the *P* spot remains clearly defined. The corresponding generator waveforms continue to exhibit identifiable voltage plateaus, although the short positive plateau develops a slight slope and its mean amplitude progressively decreases as the duty cycle is reduced. The separation between the *N* and *P* spots progressively decreases as the duty cycle is reduced. This behavior directly follows the decrease in voltage amplitude observed in the independently measured AWG waveforms. The simultaneous reduction in deflection amplitude and appearance of the weak intensity tail are therefore predominantly associated with the distortion of the waveform generated under these very short-pulse conditions rather than with the RF holder itself.

To quantify the relationship between beam intensity and residence time, the intensity of the *P* spot was normalized by the total intensity of the two spots, $\frac{I_P}{I_N+I_P}$, and plotted as a function of duty cycle for several excitation periods (Figure 8b). A linear dependence is obtained for all investigated periods: the relative intensity of the *P* spot closely reproduces the programmed duty cycle, confirming the squareness shape of the pulse. In many cases, the data points obtained for different excitation periods show excellent agreement, particularly at a 50% duty cycle. The electron intensity measured at a selected beam position can therefore be used as a direct quantitative measure of the corresponding residence time. A small deviation appears for the shortest period of 2 ns, for which the finite transit time of the beam between the two positions is no longer negligible compared with the excitation period.

The temporal limit of the complete RF holder–deflector system was finally investigated by reducing the duration of the positive-voltage pulse from the microsecond range down to 100 ps, which was the shortest pulse duration that could be reliably generated with the available AWG. Excitation conditions with duty cycles of 25% or below were selected in order to isolate the short positive pulse while providing sufficient time for the generator to recover its maximum negative level between successive pulses. For each pulse duration, the distance between the *N* and *P* spots was measured and normalized to the maximum separation of 820 nm. In parallel, the pulse amplitude delivered by the AWG was measured using the same cables and the high-bandwidth oscilloscope, and normalized to its maximum value. This comparison shown Figure 8c directly separates the reduction in deflection originating from the waveform generator from additional attenuation introduced by the holder–deflector assembly.

Down to a pulse duration of approximately 0.4 ns, the normalized beam deflection follows almost exactly the voltage decrease measured at the AWG output. The reduction in deflection observed in this regime is therefore controlled predominantly by the generator. Below 0.4 ns, the response of the holder–deflector assembly begins to contribute measurably to the attenuation. Nevertheless, after correcting for the amplitude reduction already occurring at the generator output, the deflection still reaches approximately 93% of its maximum value for a 200 ps pulse and 83% for a 100 ps pulse. These values are fully consistent with the broadband RF measurements described in Section 2 and demonstrate effective beam deflection on a 100 ps timescale. Shorter pulse durations could not be investigated with the available AWG, so the present measurements do not establish the intrinsic temporal limit of the holder–deflector assembly.

It is useful to place the present approach in the context of previously reported electrically driven RF pulser schemes. In the low-frequency "fast chopping" mode, electron pulse durations down to approximately 200 ps have been reported, with repetition rates typically around 1 MHz [27,28]. Under these conditions, short temporal windows naturally lead to low duty cycles; for example, a 2 ns temporal window at 1 MHz corresponds to a duty cycle of 0.2%. The RF-deflection approach explored here offers additional flexibility in the choice of repetition rate and temporal gate duration, allowing the duty cycle to be adjusted to the dynamics under investigation. This can be advantageous experimentally, as the fraction of the continuous electron beam contributing to the selected time window can be adapted to balance temporal resolution and collected electron statistics.

At higher repetition rates, GHz RF beam-sweeping approaches have demonstrated substantially shorter electron pulses, down to approximately 10 ps [27,28]. Such schemes rely on a more elaborate beam-manipulation architecture, including a second deflection stage to compensate for the transverse momentum imparted during the initial sweep. The present implementation does not yet reach the temporal resolution demonstrated with these GHz sweeping approaches. Instead, it explores a complementary operating regime that combines hundred-picosecond-scale beam control with flexible duty-cycle selection and a comparatively simple deflection geometry. In addition, we demonstrate compatibility with off-axis electron holography, opening the possibility of extending phase-sensitive measurements of electrostatic and magnetic fields towards time-resolved experiments.

## 4. Compatibility with high-resolution TEM and off-axis electron holography

### 4.1. Preservation of high spatial resolution

The applicability of the RF deflection approach to actual TEM experiments first requires that the pulsed-beam operation preserves the optical properties required for actual TEM experiments, and first and foremost, the spatial resolution. This was evaluated by comparing high-resolution TEM images acquired under continuous and pulsed illumination conditions (Figure 9).

A gold nanoparticle was imaged using the same conditions in both configurations. In pulsed mode, a period of 2 ns and a duty cycle of 10% were applied to the deflector, corresponding to an electron probe window of 200 ps. As expected from the duty cycle, the electron dose measured with the K3 detector was reduced by approximately a factor of ten compared with continuous illumination. Despite this substantial decrease in the number of detected electrons, no degradation of the spatial resolution was observed. As demonstrated Figure 9, the atomic-resolution contrast is preserved in the pulsed image, and comparison of the corresponding Fourier transforms reveals the same high-spatial-frequency information as under continuous illumination. The RF beam-deflection process therefore does not introduce a detectable loss of spatial resolution under these sub-nanosecond gating conditions.

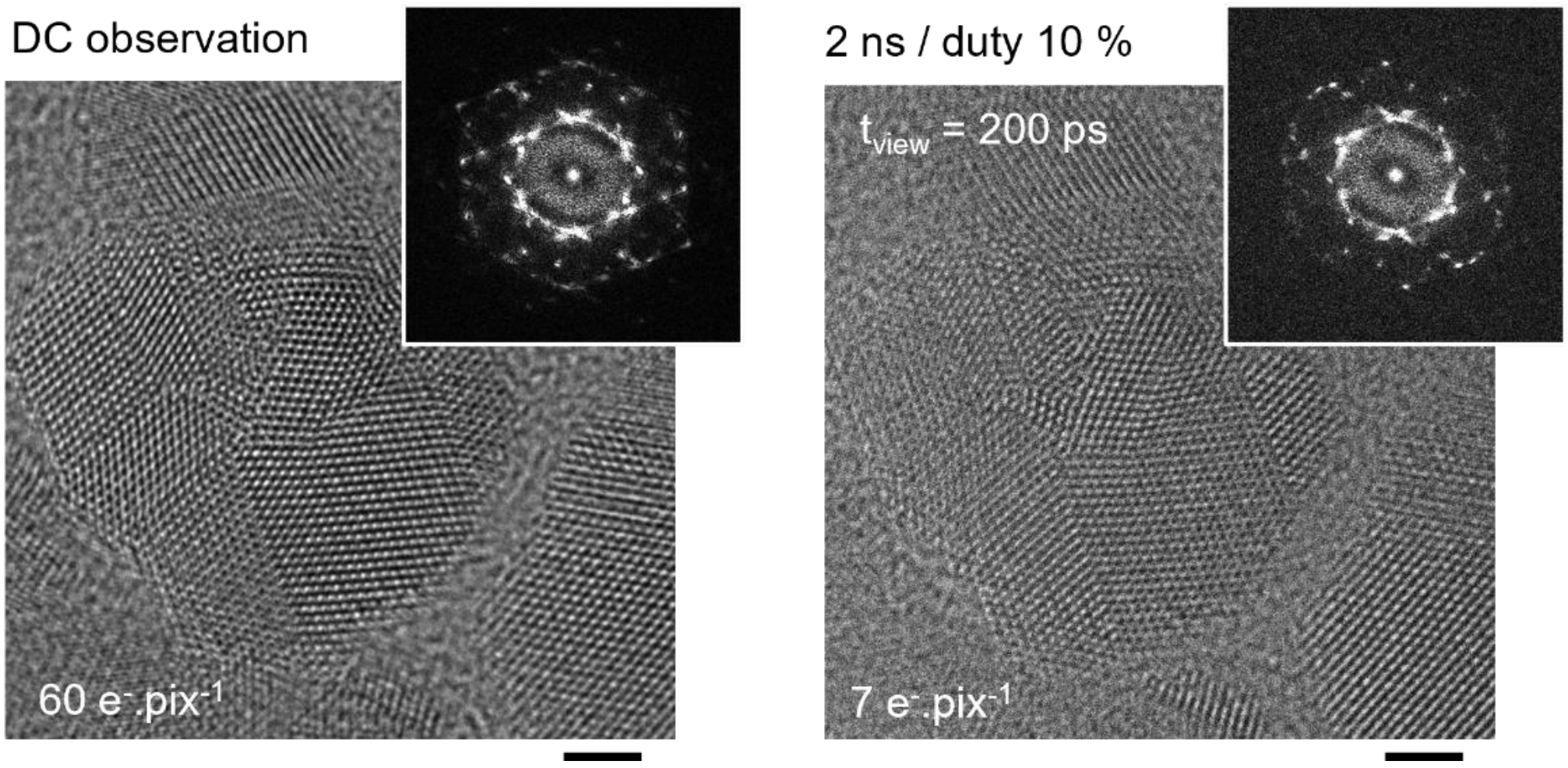

***Figure 9.*** *High-resolution TEM images of the same Au nanoparticle acquired under continuous illumination and under RF-controlled pulsed illumination with a period of 2 ns and a 10% duty cycle, corresponding to a 200 ps electron-probe window. The corresponding Fourier transforms are shown for comparison. The electron dose measured with the K3 detector is reduced by approximately a factor of ten in pulsed mode, while no detectable degradation of the spatial resolution is observed. Scale bars are 2 nm.*

This result is particularly important for a stroboscopic implementation, since the temporal selection of the electrons is achieved without modifying the electron source or introducing an additional pulsing element into the electron-optical column. The spatial performance of the microscope can therefore be retained while the effective illumination time is reduced to the sub-nanosecond range.

### 4.2. Off-axis electron holography under pulsed illumination

Off-axis electron holography provides a more demanding test of the quality of the pulsed beam because the technique relies on the spatial coherence of the electron wave and is highly sensitive to instrumental instabilities.

For these experiments, the RF deflector was driven using a Tabor SE5082 arbitrary waveform generator with a sample rate of 5 $GS \cdot s^{-1}$. Its higher output voltage, up to 1.2 V into 50 Ω, provided a beam displacement of approximately 1.8 μm with the 330 μm-wide deflector. This larger deflection allowed a field of view close to 1 μm, comparable to that previously used for *operando* electron-holography experiments on I2TEM[35–37]. The minimum pulse duration accessible with this generator was 0.5 ns.

Holograms were recorded from the same region of a specimen consisting of latex spheres deposited on a carbon membrane under three illumination conditions (Figure 9): (a) continuous illumination; (b) a 100 ns period with a 10% duty cycle corresponding to a 10 ns probe window; (c) a 5 ns period with a 10% duty cycle corresponding to a 0.5 ns probe window. All holograms were acquired using the double-biprism configuration and dynamic microscope control allowing to reach a total exposure time of 120 s. The interference-fringe spacing was 1 nm, corresponding to 6.45 pixels, and the reconstructed phase images had a spatial resolution of 2 nm.

Well-defined interference fringes are retained under both pulsed conditions, including for the 0.5 ns probe window with a visibility very close to the DC one. The reconstructed phase images remain of high quality and show the same specimen features as those obtained under continuous illumination. No additional phase distortion occurred under pulsed observations.

The pulsed holograms in Figures 9b and 9c exhibit increased noise, as expected from the tenfold reduction in collected electron dose associated with the 10% duty cycle, but this does not prevent reliable phase reconstruction. Recovering counting statistics comparable to continuous-beam operation simply requires longer acquisition times, making long-term instrumental stability increasingly important as the duty cycle is reduced. Dynamic control is particularly advantageous in this context, as it compensates specimen drift and fringe instabilities throughout extended hologram acquisitions and thereby preserves the quality of the reconstructed phase information despite the lower instantaneous electron flux [38].

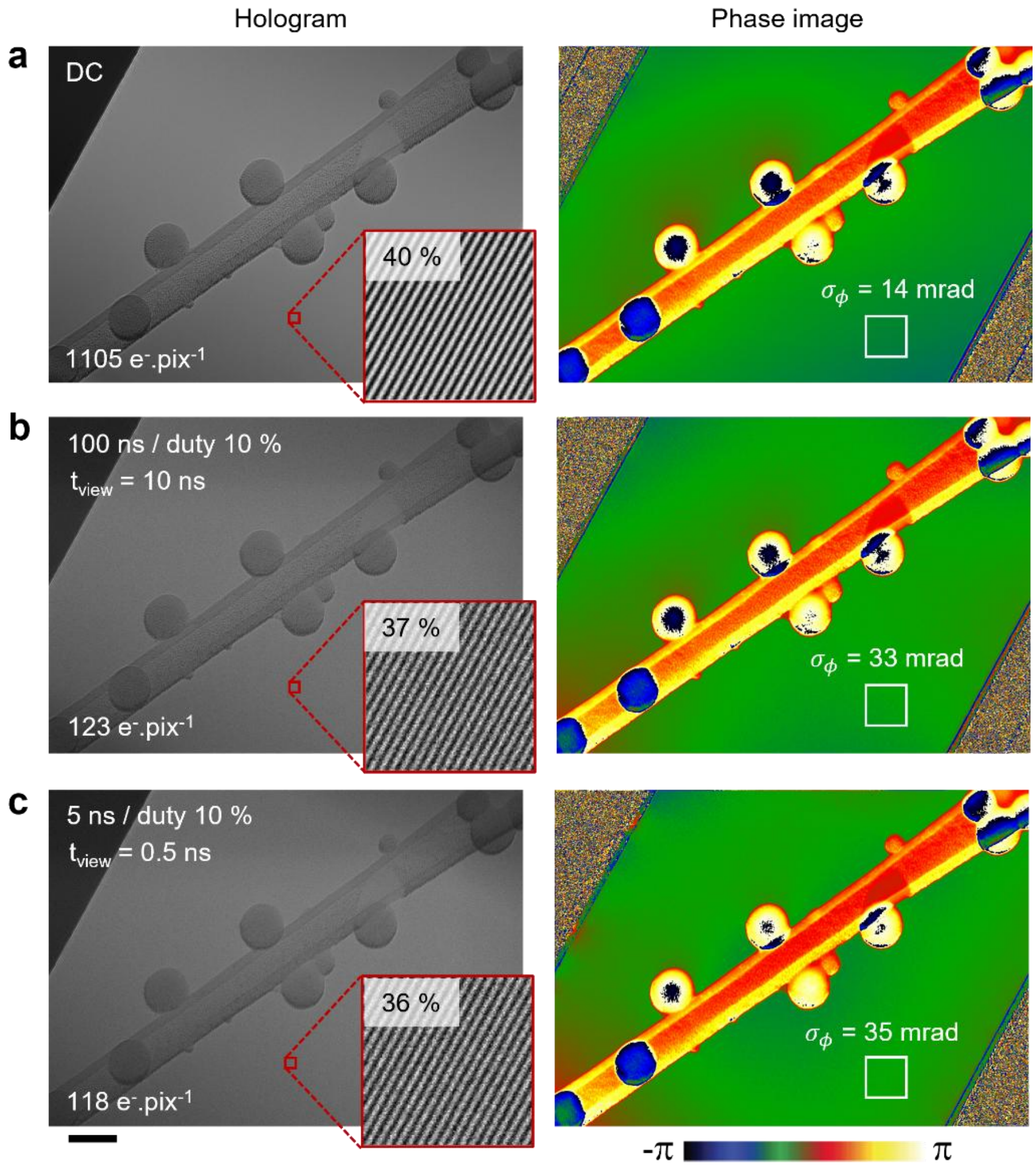

***Figure 10.*** *Electron holograms and corresponding reconstructed phase images acquired from the same region of a specimen consisting of latex spheres deposited on a carbon membrane. The insets show a detailed view of the interferometric fringes and the corresponding visibility. (a) Continuous illumination. (b) Pulsed illumination with a period of 100 ns and a 10% duty cycle, corresponding to a 10 ns probe duration. (c) Pulsed illumination with a period of 5 ns and a 10% duty cycle, corresponding to a 0.5 ns probe duration. Holograms were acquired for 120 s using dynamic control with a fringe spacing of 1 nm (6.45 pixels), yielding a spatial resolution of 2 nm in the reconstructed phase images. Scale bar is 100 nm.*

Importantly, as shown in Figure 10, temporal selection by RF beam deflection does not intrinsically reduce the fringe visibility of the transmitted electron beam compared to that obtained under continuous-beam conditions. This fundamentally differs from interference-gating approaches [39–41], in which the specimen remans continuously illuminated while interference is deliberately suppressed outside the selected temporal window, leading to a reduction in effective fringe visibility. Since the phase noise scales approximately as [42,43]

$$\sigma_{\phi} \propto \frac{1}{V\sqrt{N_e}} \qquad \text{Eq. (2)}$$

where $V$ is the fringe visibility and $N_e$ the number of collected electrons, preserving $V$ avoids this intrinsic sensitivity penalty. The reduction in $N_e$ imposed by the finite duty cycle can instead be compensated by increasing the acquisition time, particularly when combined with dynamic-control procedures that maintain fringe stability during long exposures.

These results demonstrate that the RF-deflection approach is compatible with phase-sensitive coherent imaging down to a 0.5 ns probe window, while preserving the beam coherence required for off-axis electron holography and spatial resolution. More generally, because temporal selection acts directly on the electron beam rather than on the interference process, the method is not restricted to electron holography and can in principle be extended to conventional TEM imaging, diffraction and spectroscopy, providing a versatile route toward *operando* TEM experiments in the sub-nanosecond regime.

**Conclusion**

We have developed two- and four-channel broadband RF specimen holders designed to extend *in situ* TEM experiments into the GHz and sub-nanosecond regimes. Their hybrid architecture combines low-loss coaxial transmission lines with miniaturized impedance-controlled PCB sections and removable TEM chips connected without wire bonding. VNA measurements demonstrate broadband electrical transmission through the complete RF path up to the specimen itself. In particular, the two-channel holder equipped with a representative $Si/Si_3N_4$ TEM chip exhibits a −3 dB bandwidth of 7.9 GHz, with insertion losses remaining below 12 dB up to 18 GHz.

The same RF architecture was used to develop a miniaturized electrostatic electron-beam deflector. Its measured temporal response indicates that the holder–deflector assembly remains compatible with pulse durations approaching 100 ps, although this regime is presently close to the operating limit of the available waveform generator. While the present implementation does not yet reach temporal resolutions below 100 ps, it provides broad flexibility in the choice of both pulse duration and repetition rate, allowing the duty cycle to be adapted to the dynamics under investigation and to the electron dose required for a given experiment. High-resolution TEM measurements show that sub-nanosecond operation preserves the spatial resolution of the microscope. Most importantly, off-axis electron holography performed with probe durations down to 0.5 ns demonstrates a sensitivity comparable to conventional DC operation, without measurable loss of fringe visibility under conditions representative of previously performed *operando* experiments. The approach therefore preserves the key capabilities required for quantitative electrostatic and magnetic field mapping while extending time-resolved electron holography into the sub-nanosecond regime, providing a particularly attractive route towards new advanced *operando* TEM experiments, including conventional bright- and dark-field imaging, diffraction and spectroscopy.

The present implementation is still partly limited by the available waveform generators, both in terms of the shortest pulse duration that can be reliably generated and of the voltage amplitude available for driving the deflector. Increasing the latter is particularly important for producing larger beam displacements and therefore accessing larger fields of view under normal imaging conditions. A dedicated high-bandwidth, higher-amplitude waveform generator will therefore be required to explore the ultimate temporal and deflection capabilities of the RF holder–deflector system.

The double-stage configuration of I2TEM provides a direct route towards fully electrical pump–probe TEM, with one RF holder controlling the electron probe and a second independently exciting the specimen. The remaining experimental step is to synchronize the two electrical signals and calibrate their relative delay. Because both holders have similar RF architectures and can be connected using cables of the same type and length, their electrical propagation delays should be closely matched. The principal deterministic offset will then be the electron time of flight between the two object planes: for the 48 mm separation of I2TEM at 300 keV, this corresponds to approximately 206 ps. This delay can be compensated electronically and does not constitute a limitation of the achievable temporal resolution.

Finally, the concept is not intrinsically restricted to the dual-stage geometry of I2TEM. Although the present deflector was specifically designed for the RF specimen holders, a similar combination of coaxial RF transmission, a miniaturized PCB and an electrostatic deflector could in principle be integrated at a suitable aperture-holder position in other TEM columns. Such an implementation could extend this fully electrical stroboscopic approach to more conventional microscopes without requiring major instrumental modifications. Combined with broadband electrical excitation delivered directly to the specimen, this architecture therefore opens the way to flexible, fully electrical in situ experiments from long timescales down to the sub-nanosecond regime.

## References

[1] M.L. Taheri, E.A. Stach, I. Arslan, P.A. Crozier, B.C. Kabius, T. LaGrange, A.M. Minor, S. Takeda, M. Tanase, J.B. Wagner, R. Sharma, Current status and future directions for in situ transmission electron microscopy, Ultramicroscopy 170 (2016) 86–95. https://doi.org/10.1016/j.ultramic.2016.08.007.

[2] R. Sharma, W.-C.D. Yang, Perspective and prospects of in situ transmission/scanning transmission electron microscopy, Microscopy 73 (2024) 79–100. https://doi.org/10.1093/jmicro/dfad057.

[3] J. Cumings, E. Olsson, A.K. Petford-Long, Y. Zhu, Electric and Magnetic Phenomena Studied by In Situ Transmission Electron Microscopy, MRS Bull. 33 (2008) 101–106. https://doi.org/10.1557/mrs2008.22.

[4] J.F. Einsle, C. Gatel, A. Masseboeuf, R. Cours, M.A. Bashir, M. Gubbins, R.M. Bowman, E. Snoeck, In situ electron holography of the dynamic magnetic field emanating from a hard-disk drive writer, Nano Res. 8 (2015) 1241–1249. https://doi.org/10.1007/s12274-014-0610-0.

[5] M. Brodovoi, K. Gruel, A. Masseboeuf, L. Chapuis, M. Hÿtch, F. Lorut, C. Gatel, Mapping electric fields in real nanodevices by operando electron holography, Appl. Phys. Lett. 120 (2022) 233501. https://doi.org/10.1063/5.0092019.

[6] O. Recalde-Benitez, T. Jiang, R. Winkler, Y. Ruan, A. Zintler, E. Adabifiroozjaei, A. Arzumanov, W.A. Hubbard, T. van Omme, Y. Pivak, H.H. Perez-Garza, B.C. Regan, L. Alff, P. Komissinskiy, L. Molina-Luna, Operando two-terminal devices inside a transmission electron microscope, Commun. Eng. 2 (2023) 83. https://doi.org/10.1038/s44172-023-00133-9.

[7] C. Gatel, R. Serra, K. Gruel, A. Masseboeuf, L. Chapuis, R. Cours, L. Zhang, B. Warot-Fonrose, M.J. Hÿtch, Extended Charge Layers in Metal-Oxide-Semiconductor Nanocapacitors Revealed by Operando Electron Holography, Phys. Rev. Lett. 129 (2022) 137701. https://doi.org/10.1103/PhysRevLett.129.137701.

[8] L. Zhang, F. Lorut, K. Gruel, M.J. Hÿtch, C. Gatel, Measuring Electrical Resistivity at the Nanoscale in Phase-Change Materials, Nano Lett. 24 (2024) 5913–5919. https://doi.org/10.1021/acs.nanolett.4c01462.

[9] F.J.T. Goncalves, G.W. Paterson, D. McGrouther, T. Drysdale, Y. Togawa, D.S. Schmool, R.L. Stamps, Probing microwave fields and enabling in-situ experiments in a transmission electron microscope, Sci. Rep. 7 (2017) 11064. https://doi.org/10.1038/s41598-017-11009-2.
[10] M. Möller, J.H. Gaida, S. Schäfer, C. Ropers, Few-nm tracking of current-driven magnetic vortex orbits using ultrafast Lorentz microscopy, Commun. Phys. 3 (2020) 1–7. https://doi.org/10.1038/s42005-020-0301-y.
[11] X. Fu, E. Wang, Y. Zhao, A. Liu, E. Montgomery, V.J. Gokhale, J.J. Gorman, C. Jing, J.W. Lau, Y. Zhu, Direct visualization of electromagnetic wave dynamics by laser-free ultrafast electron microscopy, Sci. Adv. 6 (2020) eabc3456. https://doi.org/10.1126/sciadv.abc3456.
[12] K.L. Ekinci, M.L. Roukes, Nanoelectromechanical systems, Rev. Sci. Instrum. 76 (2005) 061101. https://doi.org/10.1063/1.1927327.
[13] A.V. Chumak, V.I. Vasyuchka, A.A. Serga, B. Hillebrands, Magnon spintronics, Nat. Phys. 11 (2015) 453–461. https://doi.org/10.1038/nphys3347.
[14] M. Kammerer, M. Weigand, M. Curcic, M. Noske, M. Sproll, A. Vansteenkiste, B. Van Waeyenberge, H. Stoll, G. Woltersdorf, C.H. Back, G. Schuetz, Magnetic vortex core reversal by excitation of spin waves, Nat. Commun. 2 (2011) 279. https://doi.org/10.1038/ncomms1277.
[15] P. Pirro, V.I. Vasyuchka, A.A. Serga, B. Hillebrands, Advances in coherent magnonics, Nat. Rev. Mater. 6 (2021) 1114–1135. https://doi.org/10.1038/s41578-021-00332-w.
[16] M. Resano, C. Gatel, A.C. Torres-Dias, B. Lassagne, N. Biziere, Concept of imaging spin waves in pseudo-static mode by electron holography, Appl. Phys. Lett. 127 (2025) 212404. https://doi.org/10.1063/5.0294261.
[17] C. Liu, F. Ai, S. Reisbick, A. Zong, A. Pofelski, M.-G. Han, F. Camino, C. Jing, V. Lomakin, Y. Zhu, Correlated spin-wave generation and domain-wall oscillation in a topologically textured magnetic film, Nat. Mater. 24 (2025) 406–413. https://doi.org/10.1038/s41563-024-02085-7.
[18] A.H. Zewail, Four-Dimensional Electron Microscopy, Science 328 (2010) 187–193. https://doi.org/10.1126/science.1166135.
[19] T. Eggebrecht, M. Möller, J.G. Gatzmann, N. Rubiano da Silva, A. Feist, U. Martens, H. Ulrichs, M. Münzenberg, C. Ropers, S. Schäfer, Light-Induced Metastable Magnetic Texture Uncovered by in situ Lorentz Microscopy, Phys. Rev. Lett. 118 (2017) 097203. https://doi.org/10.1103/PhysRevLett.118.097203.
[20] A. Feist, N. Bach, N. Rubiano da Silva, T. Danz, M. Möller, K.E. Priebe, T. Domröse, J.G. Gatzmann, S. Rost, J. Schauss, S. Strauch, R. Bormann, M. Sivis, S. Schäfer, C. Ropers, Ultrafast transmission electron microscopy using a laser-driven field emitter: Femtosecond resolution with a high coherence electron beam, Ultramicroscopy 176 (2017) 63–73. https://doi.org/10.1016/j.ultramic.2016.12.005.
[21] A. Arbouet, G.M. Caruso, F. Houdellier, Chapter One - Ultrafast Transmission Electron Microscopy: Historical Development, Instrumentation, and Applications, in: P.W. Hawkes (Ed.), Adv. Imaging Electron Phys., Elsevier, 2018: pp. 1–72. https://doi.org/10.1016/bs.aiep.2018.06.001.
[22] A. Feist, N. Rubiano da Silva, W. Liang, C. Ropers, S. Schäfer, Nanoscale diffractive probing of strain dynamics in ultrafast transmission electron microscopy, Struct. Dyn. 5 (2018) 014302. https://doi.org/10.1063/1.5009822.
[23] T. Danz, T. Domröse, C. Ropers, Ultrafast nanoimaging of the order parameter in a structural phase transition, Science 371 (2021) 371–374. https://doi.org/10.1126/science.abd2774.
[24] T. LaGrange, P. Cattaneo, B. Barwick, D.J. Flannigan, J. Weissenrieder, F. Carbone, Laser-driven ultrafast transmission electron microscopy, Nat. Rev. Methods Primer 5 (2025) 61. https://doi.org/10.1038/s43586-025-00431-w.
[25] C. Jing, Y. Zhu, A. Liu, K. Schliep, X. Fu, Y. Zhao, E. Montgomery, W. Rush, A. Kanareykin, M. Katz, J. Lau, Tunable electron beam pulser for picoseconds stroboscopic microscopy in transmission electron microscopes, Ultramicroscopy 207 (2019) 112829. https://doi.org/10.1016/j.ultramic.2019.112829.

[26] J.W. Lau, K.B. Schliep, M.B. Katz, V.J. Gokhale, J.J. Gorman, C. Jing, A. Liu, Y. Zhao, E. Montgomery, H. Choe, W. Rush, A. Kanareykin, X. Fu, Y. Zhu, Laser-free GHz stroboscopic transmission electron microscope: Components, system integration, and practical considerations for pump–probe measurements, Rev. Sci. Instrum. 91 (2020) 021301. https://doi.org/10.1063/1.5131758.
[27] S.A. Reisbick, M.-G. Han, C. Liu, Y. Zhao, E. Montgomery, C. Jing, V.J. Gokhale, J.J. Gorman, J.W. Lau, Y. Zhu, Stroboscopic ultrafast imaging using RF strip-lines in a commercial transmission electron microscope, Ultramicroscopy 235 (2022) 113497. https://doi.org/10.1016/j.ultramic.2022.113497.
[28] S.A. Reisbick, A. Pofelski, M.-G. Han, C. Liu, E. Montgomery, C. Jing, H. Sawada, Y. Zhu, Characterization of transverse electron pulse trains using RF powered traveling wave metallic comb striplines, Ultramicroscopy 249 (2023) 113733. https://doi.org/10.1016/j.ultramic.2023.113733.
[29] A. Tonomura, Electron-holographic interference microscopy, Adv. Phys. 41 (1992) 59–103. https://doi.org/10.1080/00018739200101473.
[30] P.A. Midgley, R.E. Dunin-Borkowski, Electron tomography and holography in materials science, Nat. Mater. 8 (2009) 271–280. https://doi.org/10.1038/nmat2406.
[31] M. Lehmann, H. Lichte, Electron Holography, in: C.B. Carter, D.B. Williams (Eds.), Transm. Electron Microsc. Companion Vol., Springer Nature Switzerland, Cham, 2026: pp. 357–377. https://doi.org/10.1007/978-3-031-80789-3_12.
[32] D.M. Pozar, Microwave Engineering, 4th Edition | Wiley, Wiley, 2012. https://www.wiley.com/en-us/microwave-engineering-4th-edition-p-9781118213636 (accessed September 13, 2026).
[33] E. Snoeck, F. Houdellier, Y. Taniguch, A. Masseboeuf, C. Gatel, J. Nicolai, M. Hytch, Off-Axial Aberration Correction using a B-COR for Lorentz and HREM Modes, Microsc. Microanal. 20 (2014) 932–933. https://doi.org/10.1017/S1431927614006382.
[34] Y. Kubo, C. Gatel, E. Snoeck, F. Houdellier, Optimising electron microscopy experiment through electron optics simulation, Ultramicroscopy 175 (2017) 67–80.
[35] L. Zhang, M.H. Raza, R. Wu, K. Gruel, C. Dubourdieu, M. Hÿtch, C. Gatel, Quantification of Interfacial Charges in Multilayered Nanocapacitors by Operando Electron Holography, Adv. Mater. 37 (2025) 2413691. https://doi.org/10.1002/adma.202413691.
[36] L. Zhang, C. Gatel, M.H. Raza, K. Gruel, C. Dubourdieu, M. Hÿtch, Mapping electric fields and observation of ferroelectric domain switching in hafnia-zirconia devices by electron holography, Nat. Commun. 16 (2025) 11233. https://doi.org/10.1038/s41467-025-66807-4.
[37] B. Disic, K. Gruel, A. Masseboeuf, L. Zhang, R. Serra, M. Hÿtch, C. Gatel, Leakage-Induced Space-Charge Accumulation and Interfacial Field Redistribution in a Dielectric Nanocapacitor Revealed by Operando Electron Holography, Nano Lett. 26 (2026) 5266–5274. https://doi.org/10.1021/acs.nanolett.6c00753.
[38] C. Gatel, J. Dupuy, F. Houdellier, M.J. Hÿtch, Unlimited acquisition time in electron holography by automated feedback control of transmission electron microscope, Appl. Phys. Lett. 113 (2018) 133102. https://doi.org/10.1063/1.5050906.
[39] T. Niermann, M. Lehmann, T. Wagner, Gated interference for time-resolved electron holography, Ultramicroscopy 182 (2017) 54–61. https://doi.org/10.1016/j.ultramic.2017.06.017.
[40] T. Wagner, T. Niermann, F. Urban, M. Lehmann, Nanosecond electron holography by interference gating, Ultramicroscopy 206 (2019) 112824. https://doi.org/10.1016/j.ultramic.2019.112824.
[41] S. Gaebel, H. Çelik, D. Berger, C.T. Koch, M. Lehmann, T. Wagner, Approaching one nanosecond temporal resolution with square-wave-based control signals for interference gating, Ultramicroscopy 277 (2025) 114208. https://doi.org/10.1016/j.ultramic.2025.114208.
[42] H. Lichte, Electron holography: optimum position of the biprism in the electron microscope, Ultramicroscopy 64 (1996) 79–86. https://doi.org/10.1016/0304-3991(96)00017-4.

[43] H. Lichte, D. Geiger, A. Harscher, E. Heindl, M. Lehmann, D. Malamidis, A. Orchowski, W.-D. Rau, Artefacts in electron holography, Ultramicroscopy 64 (1996) 67–77. https://doi.org/10.1016/0304-3991(96)00018-6.

## Acknowledgements

This study has been supported through the French National Research Agency for financial support within the ANR program EHIS (ANR-21-CE42-0003), the ANR program POLARYS (ANR-23-CE42-0011) and the "Investissement d'Avenir" program reference Nos. ANR-10-EQPX-38-01 and 11-IDEX-0002, the "Conseil Regional Midi-Pyrénées" and the European FEDER-FSE for financial support within the CPER program. These results have also received funding from the European Union Horizon 2020 research and innovation programme under grant agreement No. 823717 – ESTEEM3. This work is also supported by France 2030 government investment plan managed by the French National Research Agency under grant reference PEPR SPIN – SPINCHARAC ANR-22-EXSP-0008. We would also like to sincerely thank Olivier Pascal (Laplace laboratory, Toulouse) for lending us the Tektronix AWG70001B instrument and Etienne Snoeck for his careful proofreading of the manuscript.

## Data availability

The data are available from the corresponding authors upon request.

## Declaration of competing interests

The authors declare that they have no known competing financial interests or personal relationships that could have appeared to influence the work reported in this paper.

## Appendix A. Supplementary data

No supplementary material